\documentclass{article}
\usepackage[utf8]{inputenc}
\usepackage{authblk}
\usepackage{setspace}
\usepackage[margin=1.25in]{geometry}
\usepackage{graphicx}
\graphicspath{ {./figures/} }
\usepackage{subcaption}
\usepackage{amsmath}
\usepackage[table,xcdraw]{xcolor}
\usepackage{url}
\usepackage{multirow} 
\usepackage{amssymb}
\usepackage[style=nejm, 
citestyle=numeric-comp,
sorting=none]{biblatex}
\title{A Physically Driven Long Short Term Memory Model for Estimating Snow Water Equivalent over the Continental United States.}

\author[1, 2*]{Arun M. Saranathan}
\author[1, 2]{Mahmoud Saeedimoghaddam}
\author[1, 2]{Brandon Smith}
\author[1,2]{Deepthi Raghunandan}
\author[3]{Grey Nearing}
\author[1,2]{Craig Pelissier}

\affil[1]{Science Systems and Applications, Inc., Lanham, MD, USA 20706.}
\affil[2]{NASA Goddard Space Flight Center, Greenbelt, MD, USA 20771.}
\affil[3]{Google Research, Mountainview, CA, USA 94043.}
\affil[*]{Address correspondence to: arun.saranathan@ssaihq.com}
\date{}

\begin{document}

\maketitle

%%%%%% Abstract %%%%%%
\begin{abstract}
Snow is an essential input for various land surface models. Seasonal snow estimates are available as snow water equivalent (SWE) from process-based reanalysis products or locally from \textit{in situ} measurements. While the reanalysis products are computationally expensive and available at only fixed spatial and temporal resolutions, the \textit{in situ} measurements are highly localized and sparse. To address these issues and enable the analysis of the effect of a large suite of physical, morphological, and geological conditions on the presence and amount of snow, we build a Long Short-Term Memory (LSTM) network, which is able to estimate the SWE based on time series input of the various physical/meteorological factors as well static spatial/morphological factors. Specifically, this model breaks down the SWE estimation into two separate tasks: (i) a classification task that indicates the presence/absence of snow on a specific day and (ii) a regression task that indicates the height of the SWE on a specific day in the case of snow presence. %Initially, the model was trained using SWE from the European Center for Medium-Range Weather Forecast (ECMWF) fifth-generation European Reanalysis (ERA5) products. 
The model is trained using physical/\textit{in situ} SWE measurements from the SNOw TELemetry (SNOTEL) snow pillows in the western United States. We will show that trained LSTM models have a classification accuracy of $\geq 93\%$ for the presence of snow and a coefficient of correlation of $\sim 0.9$ concerning their SWE estimates. We will also demonstrate that the models can generalize both spatially and temporally to previously unseen data.
%The abstract should be a single paragraph written in plain language that a general reader can %understand. Do not include citations, figures, tables, or undefined abbreviations in the abstract. %Any abbreviations that appear in the title should be defined in the abstract. The length should be %200 words and not exceed 250 words, to include: 
%\begin{itemize}
%    \item An opening sentence that states the question/problem addressed by the research AND
%    \item Enough background content to give context to the study AND
%    \item A brief statement of primary results AND
%    \item A short concluding sentence.
%\end{itemize} 
\end{abstract}

%%%%%% Main Text %%%%%%

\section{Introduction}

Seasonal snow represents a critical component of the hydrological cycle, serving as a key source of freshwater \cite{barnett2005climate} that regulates stream flow, groundwater recharge, and water availability for both ecological systems and anthropogenic use \cite{mankin2015snow}. Its accumulation and melt dynamics significantly influence surface energy balances, soil moisture, and vegetation processes \cite{slater2013permafrost}. In addition, seasonal snow \cite{ek2003noah} serves as an essential boundary condition in land surface models (LSM) \cite{fisher2020perspectives}, which numerically simulate the coupled fluxes of water, energy, and carbon between the land surface and the atmosphere, thus improving our ability to model and predict land-atmosphere interactions under changing climatic conditions. Recently, such LSMs have received increased attention due to the focus on building Earth System Digital Twins (ESDTs) that can provide high-fidelity, interactive digital replicas of various Earth systems \cite{balint2022digitaltwin, lee2023digitaltwin}. Snow data is available in the form of snow water equivalent (SWE) from the \textit{in situ} point measurements \cite{serreze1999snotel, brown2019canada}, which provide SWE measurements over a sparse set of stations. Estimates of snow (in the form of SWE) are also available from a variety of snow reconstruction and/or reanalysis products \cite{pulliainen2020globsnow, hersbach2020era5, margulis2016landsat, mesinger2006narr, fang2022western}. While the \textit{in situ} measurements are sparse and cannot provide detailed spatial extents of the snow-pack, the process reanalysis products are computationally expensive and require significant corrections and reanalysis to represent the observed snow covers better. Further, advanced computational models like ESDTs must have forecasting and modeling abilities for scenario planning, etc., which cannot be easily incorporated into the existing frameworks.

Machine Learning (ML)-based methods, with their ability to model even highly complex and non-linear relationships between input and output variables, have consistently demonstrated state-of-the-art performance when used as LSMs \cite{pal2021review, nearing2021role, sun2023surrogate, slater2022hybrid}. Additionally, advancements in GPU-based computing enable these models to generate predictions almost instantaneously. In the domain of SWE estimation, Pflug et al. (2025) \cite{pflug2025lightweight} have very recently introduced a lightweight LSTM model for estimating SWE. This model integrates limited meteorological forcings, such as precipitation and air temperature, with remotely sensed snow priors derived from MODIS snow cover products \cite{hall2021modis}, specifically the MOD10A1 fractional snow cover abundances ($fSCA$) aggregated to a $1~ km$ resolution. The snow product (SWE) is extracted from the Western US snow reanalysis product \cite{fang2022western}. The reanalysis products are generated by performing an ensemble of snow simulations using the SImplified Simple Biosphere LSM with the Snow-Atmosphere-Soil-Transfer (SiSB-SAST) model \cite{sun1999simplified, xue2003impact}. The manuscript reports an impressive estimation performance and the generalizing capabilities of an ML model trained in the Western US to some locations in Western Europe.

In this manuscript, we present an LSTM-based snow-specific LSM surrogate developed as part of the Terrestrial Environmental Rapid-Replication and Assimilation Hydro-meteorological (TERRAHydro) ecosystem \cite{smith2024advancing, saeedimoghaddamterrahydro}. The presented model was further built using the Coupled Reusable Earth System Tensor (CREST) technology. The proposed LSTM model uses a wide range of meteorological and physical/topographic forcing to consider the effect of a large range of climate and physiological factors on the snow-pack. Further, to enable the use of this model for forecasting and scenario modeling efforts, we do not use ``snow priors" in the form of satellite-derived $fSCA$ as input to our model. Additionally, since the actual snow cover for most of the continental US is $0$ for significant chunks of the year in most locations, the snow data has a very high sparsity. In the absence of priors such as the $fSCA$, we split the snow estimation problem into two components: a snow presence (classification) problem, and a snow depth estimation (regression) problem in the case of snow presence. The climate/meteorological time-series data used to drive these LSTM models were extracted from the European Center for Medium-Range Weather Forecast (ECMWF) fifth-generation European Reanalysis (ERA5) simulations \cite{hersbach2020era5} which provides this data at a relatively high spatial and temporal resolution, Further existing models are designed to operate indvidually, where the current model was designed and optimized to ensure compatibility and interoperability with other LSMs built as part of TERRAHydro (more information on other LSM surrogates, their interoperability etc., will be presented in an upcoming paper from Saeedimoghaddam et al. (2025)). The LSTM models are trained on snow data from physical/in situ measurements from the SNOw TELemetry (SNOTEL) snow pillows in the Western US \cite{serreze1999characteristics} provided by the Natural Resources Conservation Service. %In particular, we used the bias \cite{sun2019regional} and quality \cite{yan2018next} corrected data from these snow-pillows in our modeling efforts.

The manuscript is organized as follows: Sec. \ref{sec:data} describes the various products used in our model. Sec. \ref{sec:methods} describes the models, the architectural details, and other details of the LSTM models used in this work. Sec. \ref{sec:results} describes the results of the various experiments. Sec. \ref{sec:disc} discusses the multiple results, and finally, Sec. \ref{sec:conc} describes the conclusions and indicates some directions for future works.

\section{Data Products} \label{sec:data}

\begin{table}[]
\centering
\caption{Full List of ERA5 variables used in our snow-pack modeling. Most of these variable are used as time-series input variables, except for 'sd' which is used as the ERA5 snow-pack variable. All the variables are available at a spatial resolution of $0.25^{\circ} \times 0.25^{\circ}$.}
\label{tab:era5_var}
\begin{tabular}{|l|l|l|}
\hline
\textbf{Variable}    & \textbf{Units} & \textbf{Description} \\ \hline
\textit{cape}        &  $[J/kg]$      &  Convective Available Potential Energy             \\ \hline
\textit{cp}          &  $[m]$         &  Convective Precipitation                          \\ \hline
\textit{cvh}         &  $[]$          &  High Vegetation Cover                             \\ \hline
\textit{cvl}         &  $[]$          &  Low Vegetation Cover                              \\ \hline
\textit{d2m}         &  $[K]$         &  2 metre Dewpoint Temperature                      \\ \hline
\textit{e}           &  $[m]$         &  Evaporation                                       \\ \hline
\textit{fal}         &  $[]$          &  Forecast Albedo                                   \\ \hline
\textit{lai\_hv}     &  $[]$          &  Leaf Area Index, high Vegetation                  \\ \hline
\textit{lai\_lv}     &  $[]$          &  Leaf Area Index, low Vegetation                   \\ \hline
\textit{lsm}         &  $[]$          &  Land-Sea Mask                                     \\ \hline
\textit{msdwlwrf }   &  $[W/m^{2}]$   &  Mean Surface Downward Long-Wave radiation flux    \\ \hline
\textit{msdwswrf}    &  $[W/m^{2}]$   &  Mean Surface Downward Short-Wave radiation flux   \\ \hline
\textit{pev}         &  $[m]$         &  Potential Evaporation                             \\ \hline
\textit{skt}         &  $[K]$         &  Skin Temperature                                  \\ \hline
\textit{slhf}        &  $[J/m^{2}]$   &  Surface Latent Heat Flux                          \\ \hline
\textit{sp}          &  $[Pa]$        &  Surface Pressure                                  \\ \hline
\textit{sro}         &  $[m]$         &  Surface Runoff                                    \\ \hline
\textit{ssro}        &  $[m]$         &  Sub-Surface Runoff                                \\ \hline
\textit{ssr}         &  $[J/m^{2}]$  &  Surface net short-wave (solar) radiation          \\ \hline
\textit{ssrd}        &  $[J/m^{2}]$  &  Surface net short-wave (solar) radiation downwards \\ \hline
\textit{str}         &  $[J/m^{2}]$  &  Surface net long-wave (thermal) radiation          \\ \hline
\textit{strd}        &  $[J/m^{2}]$  &  Surface net long-wave (thermal) radiation downwards \\ \hline
\textit{stl1}        &  $[K]$         &  Soil Temperature Level-1                          \\ \hline
\textit{stl2}        &  $[K]$         &  Soil Temperature Level-2                          \\ \hline
\textit{stl3}        &  $[K]$         &  Soil Temperature Level-3                          \\ \hline
\textit{stl4}        &  $[K]$         &  Soil Temperature Level-4                          \\ \hline
\textit{swvl1}       &  $[m^{3}/m^{3}]$ & Volumetric Soil Water Layer-1                    \\ \hline
\textit{t2m}         &  $[K]$         &  2 metre Temperature                               \\ \hline
\textit{tp}          &  $[m]$         &  Total Precipitation                          \\ \hline
\textit{u10}         &  $[m/s]$       &  10 metre U-wind component                         \\ \hline
\textit{v10}         &  $[m/s]$       &  10 metre V-wind component                         \\ \hline
\textit{z}           &  $[m^{2}/s^{2}]$ &  Geopotential                                    \\ \hline
\textit{\color[HTML]{FE0000} sd}         &  {\color[HTML]{FE0000}$[m]$}         &  {\color[HTML]{FE0000} Snow Depth}                                        \\ \hline
\textit{\color[HTML]{00AC00} csfr}         &  {\color[HTML]{00AC00}$[m]$}         &  {\color[HTML]{00AC00} Convective snowfall rate water equivalent}                                        \\ \hline
\textit{\color[HTML]{00AC00} es}         &  {\color[HTML]{00AC00}$[m]$}         &  {\color[HTML]{00AC00} Snow Evaporation}                                        \\ \hline
\textit{\color[HTML]{00AC00} smlt}         &  {\color[HTML]{00AC00}$[m]$}         &  {\color[HTML]{00AC00} Snow Melt}                                        \\ \hline
\end{tabular}
\end{table}

\subsection{Model Input data}\label{subsec:input_data}
\subsubsection{Time Series Forcing Data}\label{subsec:timeSeries_vars}

Since we are proposing an LSTM-based ML model, the first requirement for our model is accurate time-series data of various physical and meteorological forcings that affect the properties of the snow-pack - considered essential inputs to a snow estimation model. To address this requirement, we leverage the European Center for Medium-Range Weather Forecast's (ECMWF) ERA5 reanalysis (\url{https://cds.climate.copernicus.eu/datasets/reanalysis-era5-single-levels?tab=overview})\cite{hersbach2020era5} products. These products combine modeled predictions with observations from various sources worldwide to create a physically consistent and globally complete dataset. ERA5 provides hourly estimates for "atmospheric, ocean-wave, and land-surface quantities" \cite{hersbach2020era5}. The land-surface and ocean-wave variables are calculated by combining the meteorological forcings with land-surface and wave models. The different quantities are available at a temporal resolution of $1~hour$ and are at a spatial resolution of $0.25^{\circ} \times 0.25^{\circ}$. Further, since physical snow data is generally available at a temporal resolution of $1~day$, we aggregate the different climatic and physical variables to this temporal resolution as well. Finally, while the data is available from $1940-$present (with a latency of ~$5$ days), we focus specifically on the time range between $2010-2021$ to better match the active time period of physical snow measurements described in Sec. \ref{subsec_snotel}.  
The full list of variables used in our modeling are shown in Table \ref{tab:era5_var}. These variables are the primary time-series inputs to our LSTM model.  Further, we specifically leverage the $'lsm'$ (Land-Sea Mask) variable to mask any locations which are not over land, as our current approach only focus on snow-pack estimation above dry land. One notable addition among the ERA5 variables is the $'sd'$ (Snow Depth variable highlighted in red in Table \ref{tab:era5_var}). Note that in spite of the slightly different naming convention, the ECMWF documentation defines the variable $'sd'$, as ``the depth the water would have if the snow melted and spread evenly over the whole grid box"\footnote{This description can be found at \url{https://codes.ecmwf.int/grib/param-db/141}.}, which is consistent with the definition of SWE as used in other datasets/publications. It should also be noted that since $'sd'$ is already a modeled estimate of the snow-pack properties from a reanalysis, it will be used as the standard against which our model is compared. Further the ERA5 database also has additional snow-related variables like Convective snowfall rate water equivalent ($'csfr'$), snow evaporation ($'es'$), and snow melt ($'smlt'$) which are also highly related to the total snow at a given place and time, these variables are highlighted in Table. \ref{tab:era5_var}. In the primary experiments in this manuscript we are only focused on building the most powerful models for estimating the snow but in Appendix \ref{app:lstm_independance} we will also show the LSTM is capable of learning the relationship between the observed snow and non-snow related meteorological forcings, that the LSTM is not only able to correct for the biases/differences present between reanalysis products and \textit{in situ} measurements, but that they are also capable of learning the relationship between the different weather variables.

\subsubsection{Static Spatial Attributes for Model Improvement} \label{subsec_staticAttributes}

% \usepackage{multirow}
\begin{table}[]
\caption{The full list of static spatial variables used in our snow-pack model.}
\label{tab:static_spatial_var}
\begin{tabular}{|l|l|l|l|}
\hline
\textbf{Dataset} & \textbf{\begin{tabular}[c]{@{}l@{}}Spatial \\ Resolution\end{tabular}} & \textbf{Variables} & \textbf{Description} \\ \hline
\multirow{3}{*}{\textit{\textbf{Soil}}} & \multirow{3}{*}{$0.25^{\circ} \times 0.25^{\circ}$} & \textit{clay\_frac} & Fraction of clay. \\ \cline{3-4} 
 &  & \textit{sand\_frac} & Fraction of sand. \\ \cline{3-4} 
 &  & \textit{silt\_frac} & Fraction of silt. \\ \hline
\multirow{5}{*}{\textit{\textbf{Irrigation}}} & \multirow{5}{*}{$0.25^{\circ} \times 0.25^{\circ}$} & \textit{Actual} & Area actually irrigated. \\ \cline{3-4} 
 &  & \textit{Equipped} & Area equipped for irrigation. \\ \cline{3-4} 
 &  & \textit{Ground water} & \begin{tabular}[c]{@{}l@{}}Area equipped for irrigation\\  with ground water.\end{tabular} \\ \cline{3-4} 
 &  & \textit{Non-Conventional} & \begin{tabular}[c]{@{}l@{}}Area equipped for irrigation \\ with non-conventional sources\end{tabular} \\ \cline{3-4} 
 &  & \textit{Surface waters} & \begin{tabular}[c]{@{}l@{}}Area equipped for irrigation\\ with surface waters.\end{tabular} \\ \hline
\textit{\textbf{Topography}} & $0.075^{\circ} \times 0.075^{\circ}$ & \textit{elev} & \begin{tabular}[c]{@{}l@{}}Void filled Digital Elevation \\ Model.\end{tabular} \\ \hline
\textit{\textbf{\begin{tabular}[c]{@{}l@{}}MODIS Landcover\\ (MCD12C1)\end{tabular}}} & $0.2^{\circ} \times 0.2^{\circ}$ & \textit{\begin{tabular}[c]{@{}l@{}}Majority\_Land\\ \_cover\_Type1\end{tabular}} & \begin{tabular}[c]{@{}l@{}}Most Likely land-cover class \\ based on the IBGP model.\end{tabular} \\ \hline
\multirow{7}{*}{\textit{\textbf{\begin{tabular}[c]{@{}l@{}}MODIS Vegetation\\ Indicies (MOD13C2)\end{tabular}}}} & \multirow{7}{*}{$0.2^{\circ} \times 0.2^{\circ}$} & \textit{Monthly EVI} & \begin{tabular}[c]{@{}l@{}}The calculated Enhanced \\ Vegetation Index (EVI).\end{tabular} \\ \cline{3-4} 
 &  & \textit{Monthly NDVI} & \begin{tabular}[c]{@{}l@{}}The calculated Normalized\\ Difference Vegetation Index\\  (NDVI).\end{tabular} \\ \cline{3-4} 
 &  & \textit{\begin{tabular}[c]{@{}l@{}}Monthly MIR\\  reflectance\end{tabular}} & \begin{tabular}[c]{@{}l@{}}The surface reflectance of the \\ MODIS Middle InfraRed \\ (MIR: Band 7)\end{tabular} \\ \cline{3-4} 
 &  & \textit{\begin{tabular}[c]{@{}l@{}}Monthly NIR \\ reflectance\end{tabular}} & \begin{tabular}[c]{@{}l@{}}The surface reflectance of the \\ MODIS Near InfraRed \\ (NIR: Band 2)\end{tabular} \\ \cline{3-4} 
 &  & \textit{\begin{tabular}[c]{@{}l@{}}Monthly blue\\ reflectance\end{tabular}} & \begin{tabular}[c]{@{}l@{}}The surface reflectance of the\\ MODIS blue band\\ ( Band 3)\end{tabular} \\ \cline{3-4} 
 &  & \textit{\begin{tabular}[c]{@{}l@{}}Monthly red\\ reflectance\end{tabular}} & \begin{tabular}[c]{@{}l@{}}The surface reflectance of the\\ MODIS red band\\ (Band 1)\end{tabular} \\ \cline{3-4} 
 &  & \textit{\begin{tabular}[c]{@{}l@{}}Monthly pixel\\ reliability\end{tabular}} & \begin{tabular}[c]{@{}l@{}}Quality reliability of the\\  Vegetation Index pixels.\end{tabular} \\ \hline
\end{tabular}
\end{table}

In addition to the time-series forcings variables, our LSTM model also leverages a set of static-in-time physio-morphological variables as inputs. These variables are crucial for informing the model about the physical and morphological characteristics of the land surface, which influence processes such as snow-pack accumulation and ablation.

The first static dataset is the Global Land Data Assimilation System (GLDAS) dataset (\url{https://ldas.gsfc.nasa.gov/gldas/soils}) \cite{rodell2004global}, which provides information on soil composition. Specifically, the model utilizes the fractions of clay, sand, and silt, resampled to a spatial resolution of $0.25^{\circ} \times 0.25^{\circ}$, to represent the surface texture and soil properties.

The second static dataset is version-5 of the Global Map of Irrigation Areas (\url{https://www.fao.org/aquastat/en/geospatial-information/global-maps-irrigated-areas/latest-version}) \cite{siebert2007global} which provides globally distributed data on irrigated areas and irrigation types. The irrigation data is available at a spatial resolution of $0.083^{\circ} \times 0.083^{\circ}$, but we reduce the spatial resolution by a factor of $3$ to match the spatial resolution of the other datasets available, i.e., $0.25^{\circ} \times 0.25^{\circ}$. 

The third static dataset included in our models in the Void-Filled Digital Elevations Models from the HydroSHEDS dataset (\url{https://www.hydrosheds.org/hydrosheds-core-downloads}) \cite{wickel2007hydrosheds}. The topographic data is available at a resolution of $0.0083^{\circ} \times 0.0083^{\circ}$. Since all the other datasets have a significantly coarser resolution and to enable faster processing, we downsample this data by a factor of $9$, resulting in a resolution of $0.075^{\circ} \times 0.075^{\circ}$.

The fourth static dataset is a land cover product. In our work, we extract the remote sensing-based land-cover product at the $0.05^{\circ} \times 0.05^{\circ}$ Climate Modeling Grid (CMG) from the MODerate Resolution Imaging Spectrometer (MODIS) combined Terra/Aqua yearly Land Cover product (MCD12C1) (\url{https://lpdaac.usgs.gov/products/mcd12c1v061/}) \cite{friedl2015mcd12c1}. This product includes land cover classification, land-cover percentage, as well as the model's confidence in its predictions from three sources; namely, the International Geosphere-Biosphere Programme (IGBP) model, second from University of Maryland (UMD) model, and finally from the MODIS-derived Leaf Area of Index (LAI) scheme. Since we only attempt to provide our model with information on the land-cover class at a specific location, we only use the IGBP model's land-cover predictions as input, classifying each pixel into one of $16$ possible land-cover classes. This data is available at a yearly resolution. 

The final static dataset is the Terra Moderate Resolution Imaging Spectroradiometer (MODIS) Vegetation Indices Monthly (MOD13C2) Version 6.1, which provides the vegetation indices in the form of Normalized Difference Vegetation Index (NDVI), as well as the Enhanced Vegetation Index. In addition to these values, we also include the reflectance of the Near- \& Mid-Infra Red, the blue, the red, and the pixel reliability. The dataset is also available for the CMG spatial grid mentioned above. For ease of handling, we downsample the two MODIS spatial products along each direction to generate products at a resolution of $0.2^{\circ} \times 0.2^{\circ}$.

\begin{figure}[h]
    \centering
    \includegraphics[width=\textwidth, height=4.0in]{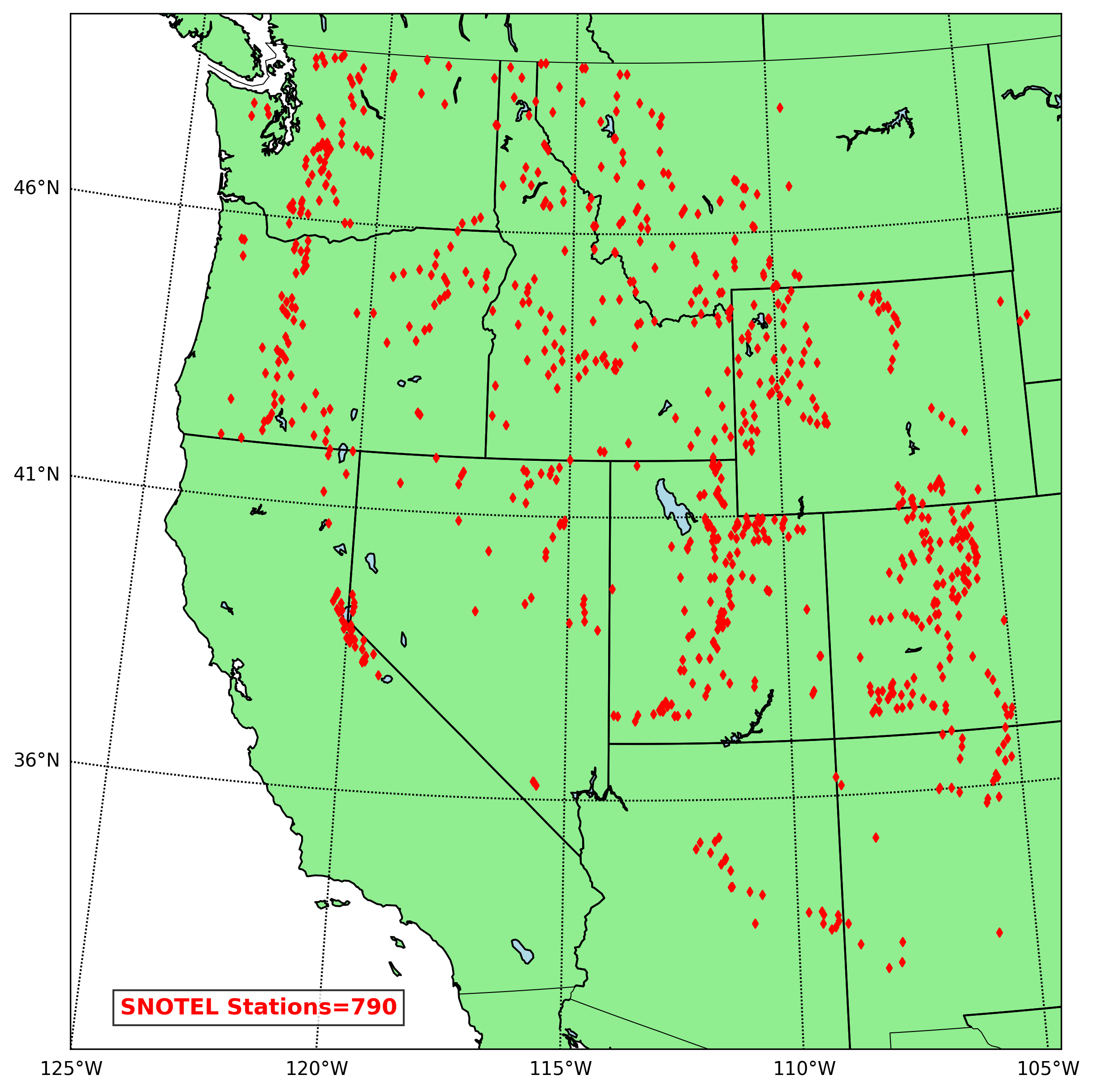} % Change this to your image file name (without extension)
    \caption{The distribution of SNOTEL \cite{serreze1999snotel} snow-pillow towers across the Western United States.}
    \label{fig:snotel_snow_towers}
\end{figure}

\subsection{Model Output data}\label{subsec:output_data}
\subsubsection{SNOTEL: \textit{In Situ} snow-pack measurements} \label{subsec_snotel}
The main source of \textit{in situ} measurements of SWE records are the 829 active SNOw TELemetry (SNOTEL) snow pillows/towers in the western US and Alaska operated by the US Department of Agriculture (USDA) Natural Resources Conservation Service (NRCS). The raw data measured at these towers is subjected to a three-stage quality control process \cite{yan2018next} that identifies and eliminates (i) outliers and errors (like -ve SWE values); (ii) extreme statistical outliers; and (iii) inconsistent SWE and precipitation values. Following this initial quality control (QC) process, the data is subjected to a bias correction in temperature and precipitation \cite{sun2019regional}. The bias-corrected and quality-controlled (BCQC) SNOTEL data (\url{https://www.pnnl.gov/data-products}) contains daily SWE, meteorological records (in the form of daily minimum, maximum, and mean air temperature), and daily total precipitation through $09/30/2021$ for all the active stations. Given our focus on CONUS, we only use the data corresponding to the $790$ stations in the continental US (the other $39$ towers are in Alaska). The spatial distribution of the various SNOTEL stations considered in this study is shown in Fig. \ref{fig:snotel_snow_towers}. The SWE measurements are stored in a labeled $'swe'$.

\section{Building the LSTM-based surrogate for estimating snow-pack variables} \label{sec:methods}
 
\subsection{Data Preprocessing \& Management} \label{subsec:dataPreproc}
For our current model, the full set of inputs are the variables shown in Tables \ref{tab:era5_var} \& \ref{tab:static_spatial_var}. The first step was to address the ERA5 variables, which include a variety of meteorological forcings and measurements of the conditions in each location (most of which are continuous physical quantities). First, to prepare these variables for use with a neural network based model, an Interquartile Range (IQR) scaling standardizes the data by subtracting the median and dividing by the IQR \cite{data2016exploratory}. Following this, we further rescale the values generated by the IQR scaling to be in the $[-1, 1]$ range. Specific ERA5 variables, namely, $['cvl', 'cvh', 'fal', 'swvl1']$ (see Table. \ref{tab:era5_var} for more details) which are either flags or are already in the required range, are excluded from these scaling operations. Among the static variables, the soil variables are fractions and therefore already in the range $[0, 1]$ and so are left as is. The irrigation variables are converted from percentages to fractions. The land-cover classes are converted from integer flags to vectors using a one-hot encoding scheme. The MODIS vegetation data are rescaled and thresholded using the valid ranges and scaling factors provided with the data \footnote{see the "Layers/Variables" section on the source page for these factors}.

An important point to note for the various datasets mentioned in Secs. \ref{subsec:timeSeries_vars} and \ref{subsec_staticAttributes} is that they are available for locations/stations on a $2D$ spatial grid. One issue is that each of the different datasets use their own spatial grids with differing spatial resolutions; further, the SNOTEL dataset (described in Sec. \ref{subsec_snotel}) is only available for selected locations or an irregular $2D$ grid, unlike the other datasets. Due to these differences, matching the samples from one grid/graph with the others is necessary. As a first step, we split all the data into equally sized spatial and temporal blocks that are overlapped such that matches between blocks are still available (for the static data described in Sec. \ref{subsec_staticAttributes}, the split is only spatial as there is no temporal dimension to these datasets). Following this, we match a given variable with specific grid locations in each spatiotemporal split with the other variables in the same block from grid locations that are closest spatially. Further, to account for incomplete/irregular grids like the SNOTEL towers, where the tower locations are irregularly distributed (see Fig. \ref{fig:snotel_snow_towers} for the exact location of the SNOTEL towers), we also place proximity constraints (both spatial and temporal) between the grid positions. A sample is considered as ``selected" for processing with our model if it can generate valid (non-null) values for all the variables via the matching scheme described here. Further, attempts are made to ensure that each batch is ``balanced" by extracting a similar number of samples from each spatiotemporal block/split in the training data.

In our implementation, we used snow data between the dates $01/01/2011~-~12/31/2021$ for model training and testing (while snow data starts from $01/01/2011$, due to the $1$ year look-back we also include meteorological data from $2010$). We further split the training data into $64$ spatio-temporal blocks ($8$ splits along both longitude and latitude, each block contains data from the entire training or validation date-range). Similarly the test data is also split into $64$ spatiotemporal blocks. Preliminary data analysis revealed that in spite of the choice of locations with high likelihood of snow, the SNOTEL $'swe'$ data is pretty sparse, with $\sim 50\%$ of the samples have an $'swe'$ of $0m$. To address issues with the expected issues with measuring very low snow values consistently due todifferences in local conditions, vegetation/plant life, and other factors, we choose to exclude measures of snow below $0.01m$ or $1 cm$ from our analysis. In addition to thresholding the $'swe'$ variable during pre-processing, a new variable $'swe\_class'$ is also added to the SNOTEL dataset according to Eq. \ref{eqn:snow_thresh}. This variable functions as a classification label indicating whether there was snow on a given day.
\begin{equation}
'swe\_class' =
\begin{cases}
    1 & \text{if } 'swe' > 0.01 \\
    0 & \text{if } 'swe' \leq 0.01
\end{cases}
\label{eqn:snow_thresh}
\end{equation}

\begin{figure}[ht]
    \centering
    \includegraphics[width=\textwidth]{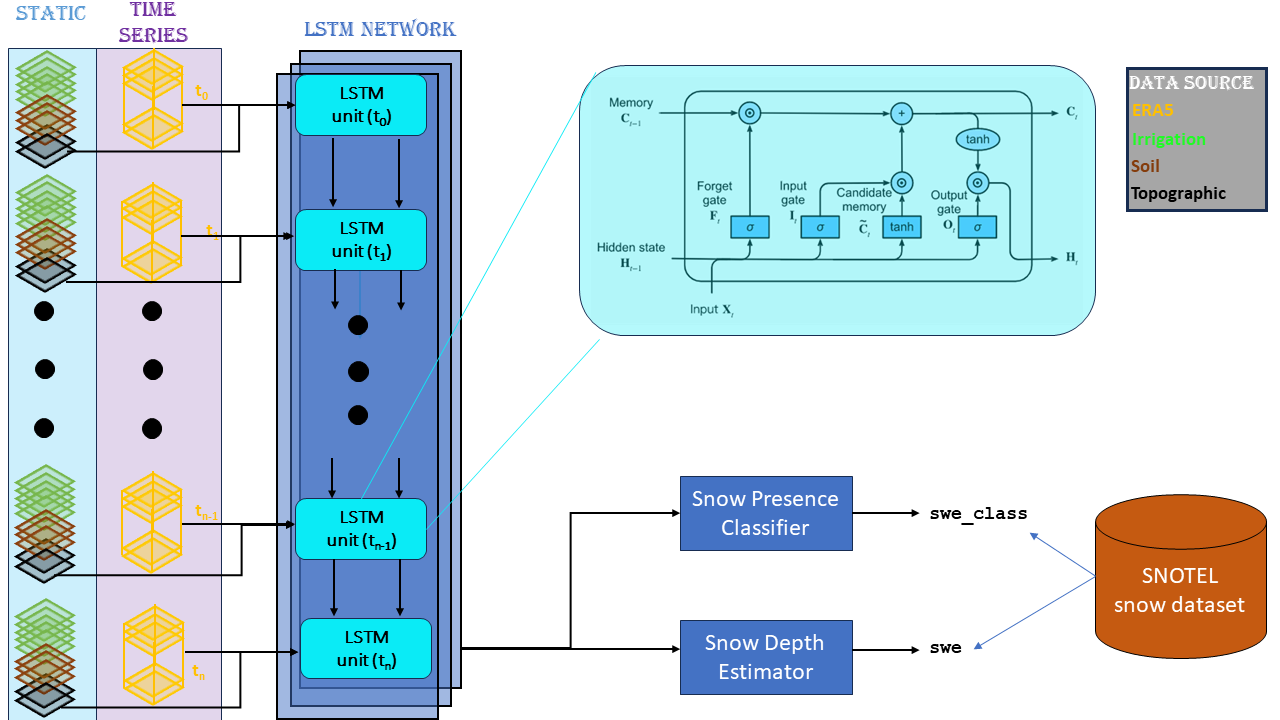} % Change this to your image file name (without extension)
    \caption{The architecture of the LSTM based model used to estimate the snow pack properties based on meteorological and physio-morphological data.}
    \label{fig:crest_terrahydro_snotel_DT}
\end{figure}

\subsection{Model Architecture and Training Details} \label{subsec:modelDetails}
Following the descriptions in Sec. \ref{subsec:dataPreproc}, we are able to generate valid samples from both the snow and non-snow datasets such that all variables in input and output are non-null. The core ML-component of our model is a NN-based architecture that is shown in Fig. \ref{fig:crest_terrahydro_snotel_DT}. The proposed LSTM model has a lookback period of $365$ time-steps. The reduced temporal resolution of $1$ day for the ERA5 variables (described in Sec. \ref{subsec:timeSeries_vars}) leads to the lookback period corresponding to $365$ days or $1$ year. Since we have both static and dynamic variables, we use a \emph{static-repetition} scheme \cite{lin2018early, rahman2020predicting} wherein we feed in the static data along with time-series input at the specific step to each LSTM cell. The set of variables are provided as input to a layer with $256$ LSTM nodes. The output of these LSTM cells are provided as the input to both the classification and regression modules. The classification module is a single dense layer that connects the LSTM output to a single node with sigmoid activation. The regression module has two dense layers, the first with $128$ nodes and the second with $1$ node. Further, we update the $'swe'$ prediction for each test sample by multiplying it with the rounded classification output in order to ensure there is $0~'swe'$ predictions for days with no snow.

In this implementation, a separate loss function is used to train the classifier and the regressor. The classifier uses a simple binary cross-entropy loss (defined as $\mathcal{L}_{\text{class}}$ in Eqn. \ref{eqn:lstm_model_loss}). The regressor uses a zero-masked Root Mean Squared Error (zRMSE) (defined as $\mathcal{L}_{\text{reg}}$ in Eqn. \ref{eqn:lstm_model_loss}). The full model loss ($\mathcal{L}$) function is defined as a weighted combination of the two as shown below.

\begin{equation}
\begin{array}{c}
    \mathcal{L}_{\text{class}} = -\frac{1}{n} \sum_{i=1}^{n} \left[ swe\_class_i \log(\widehat{\text{swe\_class}}_i) + (1 - swe\_class_i) \log(1 - \widehat{\text{swe\_class}}_{i}) \right] \\
    \mathcal{L}_{\text{reg}} = \dfrac{\sum_{i=1}^{n} (\text{swe} - \widehat{\text{swe}})_i^2 \cdot \mathbb{I}(\text{swe}_i \neq 0)}{\sum_{i=1}^{n} \mathbb{I}(\text{swe}_i \neq 0)} \\
    \mathcal{L}_{\text{model}} = \mathcal{L}_{\text{reg}} + 0.1 \mathcal{L}_{\text{class}}
\end{array}
\label{eqn:lstm_model_loss}
\end{equation}

The models were trained using an Adam optimizer \cite{kingma2014adam}, with an initial learning rate of $0.001$ and an exponential decay at the end of each epoch (with a decay rate of $0.92$). The model is trained for $50$ epochs, with a total of $250$ steps-per-epoch. 

\subsection{Model Testing \& Performance comparison experiments} \label{subsec:modelTesting}
The model's performance and accuracy are evaluated for both 'snow-presence' (classification) and 'snow-depth' (regression) under various scenarios to get a well-rounded perspective on model performance. These scenarios include: (i) Comparison of the LSTM $'swe'$ and the ERA5 $'sd'$ on a temporal test set of the \textit{in situ} measurements; (ii) Temporal cross-validation; and (iii) Spatio-temporal cross-validation. The exact details of these three testing methods are described below.
\subsubsection{Comparison of the LSTM $'swe'$ estimates to ERA5 $'sd'$ on temporal test set} \label{ss_exp1_tempTest}

The first experiment evaluates the performance of the two components of the LSTM model using a classical test set scenario, where the model is tested on previously unseen data. To accomplish this, we train the LSTM model using various static variables as well as climate variable sequences from a period spanning $01/01/2010$, to $12/31/2019$. The test set for this experiment consists of samples from $01/01/2021 - 12/31/2021$. In this experiment, we deliberately exclude samples from $01/01/2020 - 12/31/2020$ because our models have a one-year lookback period. Including $2020$ would introduce scenarios where the input data in the testing period overlaps with the training period, potentially contaminating the test set. To prevent temporal information leakage, we introduce a buffer period between the training and test sets.

The two components of the LSTM model are then tested separately and compared with the ERA5 $'sd'$ variable. The 'snow-presence' (classification) module is assessed based on how accurately it predicts the presence/absence of snow on a given day at a specific location. Likewise, we estimate whether the ERA5 model predicts significant snow accumulation (i.e. $\geq 0.01m$) on that day at that location. These predictions are evaluated against actual \textit{in situ} snow measurements from various SNOTEL towers. Similarly, the regression module of the LSTM (which estimates $'swe'$) and ERA5 (in terms of the variable $'sd'$) are evaluated by measuring the difference between their predictions and the \textit{in situ} SNOTEL measurements. The specific metrics used for evaluating classification and regression performance, along with the rationale behind their selection, are discussed in Section \ref{subsubsec:perfMetrics}.

\subsubsection{Temporal Cross Validation}\label{ss_exp2_tempCV}
While the first experiment described above provides insight into the LSTM model's ability to generalize to unseen test samples, it is important to note that snowfall and accumulation patterns vary significantly across different years and locations due to local conditions. Therefore, the second experiment is designed to evaluate how the LSTM model’s performance changes when tested on different years. We refer  to this approach of holding out  an specific year for testing in each fold as \emph{temporal cross-validation}. In each fold of this cross-validation, we also mask out the year before and after the chosen test year to prevent information leakage. For example, in the fold where the test samples range from $01/01/2015 - 12/31/2015$, the training set consists of data from $[01/01/2010-12/31/2013, 01/01/2016-12/31/2021]$. Similarly, we train $11$ LSTM models which consider samples from a specific year in the range $2011-2021$ as the test samples.  [N.B.: We do not train a model with 2010 as the test year, since this would require climate data from 2009, which is not included in our dataset.] 

The performance of the model in each fold is evaluated for both the snow-presence (classification) and snow-depth (regression) components. Further, the performance of the LSTM predictions is compared with the ERA5 estimate performance over the same period. This experiment is designed to provide key insights into the LSTM model, specifically: (i) ensuring that the improved estimation performance of the LSTM model remains consistent across different test samples; (ii) understanding the variability in LSTM performance when applied to different test sets; (iii) identifying "outlier" time periods where model performance deviates significantly and analyzing the causes of these extreme effects. It should also be noted in terms of the cross validation (both in this and the next experiment) that it is only performed for the LSTM the ERA5 estimates are only used to provide a measuring stick against which the LSTM performance can be measured.

\begin{figure}[h]
    \centering
    \includegraphics[width=\textwidth, height=4.0in]{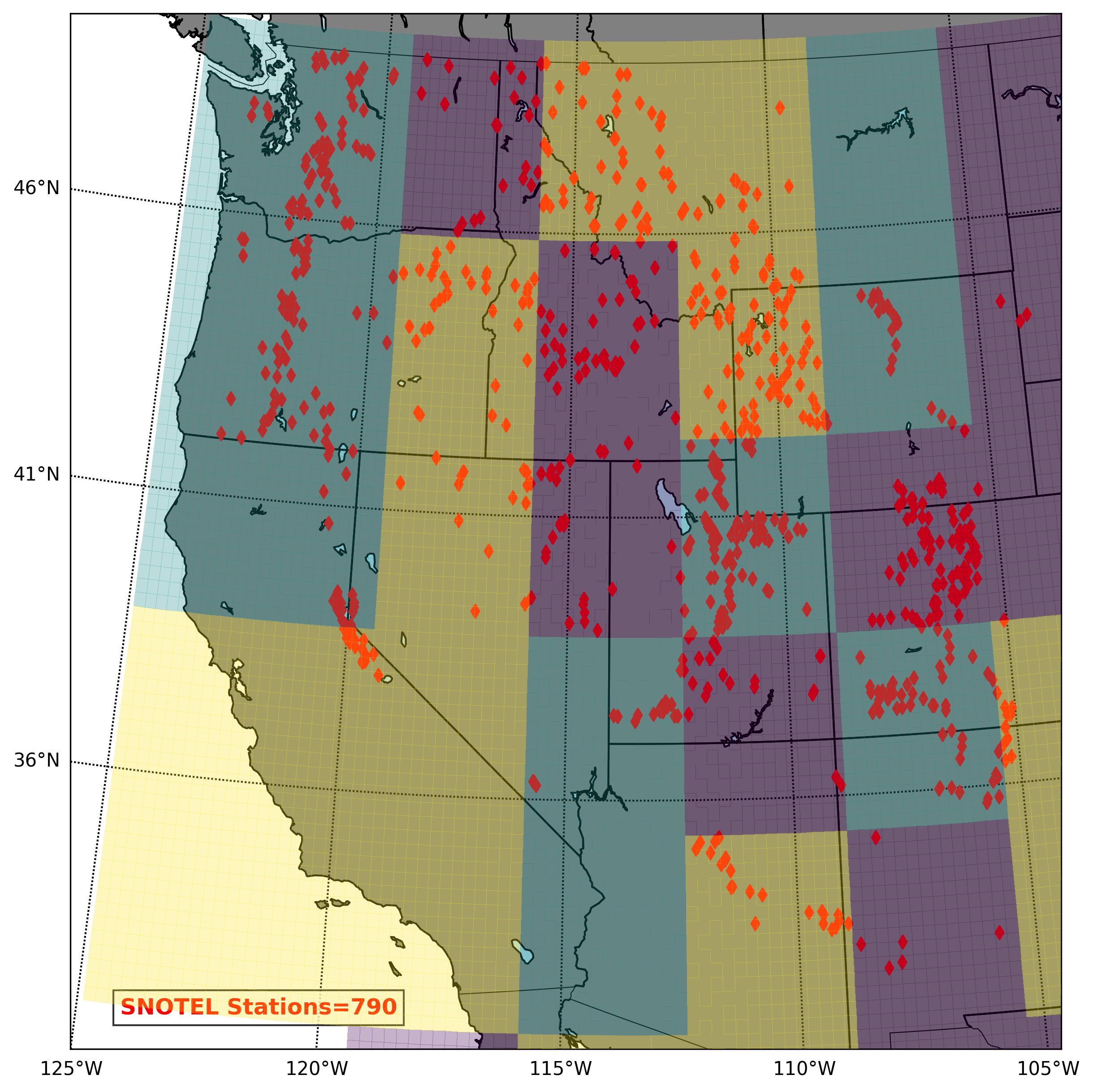} % Change this to your image file name (without extension)
    \caption{The distribution of SNOTEL \cite{serreze1999snotel} snow-pillow towers across the various spatial splits used in experiment in Sec. \ref{ss_exp3_spatTempCV}}
    \label{fig:snotel_spatTempCV_snow_towers}
\end{figure}

\subsubsection{Spatio-Temporal Cross Validation}\label{ss_exp3_spatTempCV}

The second experiment takes significant steps toward evaluating the ability of the LSTM-based model to generalize to previously unseen test samples, particularly over the temporal range. However, it is important to note that the training data — which consists of locations corresponding to known SNOTEL towers — is somewhat limited. As a result, the previous tests do not provide insights into the model’s ability to generalize to completely new test regions. To address this limitation, the final experiment is designed to provide a preliminary assessment of how well the models can generalize to both unseen spatial regions and unseen temporal ranges. To achieve this, the final experiment is structured as a spatio-temporal cross-validation. First, we split the Western US region into three spatial segments (splits). These spatial splits were generated randomly, but efforts were made to ensure an approximately equal number of SNOTEL towers in each split, as shown in Fig. \ref{fig:snotel_spatTempCV_snow_towers}. Next, we split the data temporally into 11 segments, following the same approach as in the previous experiment, resulting in a total of 33 individual models. For cross-validation, each model is trained only on data from the other two spatial splits and non-overlapping temporal splits. For example, consider a model where the test samples come from Spatial Split-1 and the year $2015$. In this case, the model is trained only on samples from Spatial Split-2 and Spatial Split-3, using data from the temporal range [$01/01/2010–12/31/2013$, $01/01/2016–12/31/2021$]. This ensures that each model is tested on data from both a spatial region and a temporal range that were completely unseen during training.

\subsubsection{Performance evaluation metrics} \label{subsubsec:perfMetrics}

For each of the experiments described above, we require a comprehensive set of metrics to evaluate the performance of the LSTM outputs and compare them with the ERA5 estimates. The 'snow-presence' (classification) predictions of the LSTM represent a classical binary classification problem, where the model predicts whether snow was present at a specific location on a given day. Accordingly, the primary metric for this analysis is the classification accuracy of the LSTM $swe\_class$ predictions, and the thresholded ERA5 $sd$ estimates. In addition to classification accuracy, we also report precision and recall, as these metrics help identify whether the LSTM or ERA5 models are prone to specific types of errors on the test set. Ideally, all three metrics would be 1 (or as close to it as possible), indicating perfect classification performance. The mathematical formulation and detailed description of these metrics are provided in Equation \ref{eqn:class_metrics}.

\begin{equation}
\begin{aligned}
\text{Accuracy} &= \frac{\text{TP} + \text{TN}}{\text{TP} + \text{TN} + \text{FP} + \text{FN}} \\
\text{Precision} &= \frac{\text{TP}}{\text{TP} + \text{FP}} \\
\text{Recall} &= \frac{\text{TP}}{\text{TP} + \text{FN}}
\end{aligned}
\label{eqn:class_metrics}
\end{equation}

\noindent
where:
\begin{itemize}
    \item \textbf{TP} = True Positives (correctly predicted days with snow)
    \item \textbf{TN} = True Negatives (correctly predicted days with no snow)
    \item \textbf{FP} = False Positives (incorrectly predicted days with snow)
    \item \textbf{FN} = False Negatives (incorrectly predicted days with no snow)
\end{itemize}

For the 'snow-depth' (regression) problem, we use a suite of well-known regression metrics to provide a comprehensive assessment of the LSTM’s performance in comparison to the ERA5 estimates. The first metric chosen is the Mean Absolute Error (MAE), which quantifies the average prediction error for each sample in the test set. Second, we consider the Median Absolute Error (MdAE), which represents the median error among the test samples. This metric is included to ensure that isolated outliers with extreme errors are detected and flagged. The third metric used in our analysis is prediction bias (Bias), which provides insights into whether the model exhibits a systematic under-prediction or over-prediction. Ideally, for MAE, MdAE, and Bias, the optimal value should be 0. The final regression metric considered in this study is the Pearson correlation coefficient (Pearson-r), which measures the overall correlation between predicted and actual values. This metric helps verify whether the model consistently captures trends such as increases and decreases in the target variable. The ideal value in this case $1$. In concert, the metrics should provide insight into the types of errors seen for the specific models. The mathematical formulation for these regression metrics is provided in Equation \ref{eqn:reg_metrics}.

\begin{equation}
\begin{aligned}
\text{MAE} &= \frac{1}{n} \sum_{i=1}^{n} \left| \text{swe}_i - \widehat{\text{swe}}_i \right| \\
\text{MdAE} &= \text{median} \left( \left| \text{swe}_i - \widehat{\text{swe}}_i \right| \right) \\
r &= \frac{\sum_{i=1}^{n} (\text{swe}_i - \bar{\text{swe}}) (\widehat{\text{swe}}_i - \bar{\widehat{\text{swe}}})} 
{\sqrt{\sum_{i=1}^{n} (\text{swe}_i - \bar{\text{swe}})^2} \sqrt{\sum_{i=1}^{n} (\widehat{\text{swe}} - \bar{\widehat{\text{swe}}})^2}} \\
\text{Bias} &= \frac{1}{n} \sum_{i=1}^{n} \left( \widehat{\text{swe}}_i - \text{swe}_i \right)
\end{aligned}
\label{eqn:reg_metrics}
\end{equation}

\subsubsection{ERA5 vs LSTM comparison} \label{ss_exp4_snotelStatPerf}

\textbf{A. Station-level comparison}: The final set of experiments involves a direct comparison of the ERA5 \texttt{sd} variable and the LSTM-estimated $'swe'$ variable (masked appropriately by $'swe\_class'$) against the \textit{in situ} measurements recorded by the SNOTEL towers. This analysis aims to assess the LSTM's ability to accurately capture the various snow dynamics observed at tower locations. Here, we use the term "snow dynamics" to refer to temporal changes in the snow-pack, including onset, accumulation, peak, ablation, and complete melting.

For each station with valid snow information throughout the entire analysis period from \texttt{01/01/2010} to \texttt{12/31/2021}, we compute two metrics of predictive accuracy:
\begin{itemize}
    \item \textit{Difference in Snow Days per Year (days)}: The median difference in the number of snow days per year between the ERA5/LSTM predictions and the observations recorded by the SNOTEL towers. This metric reflects the ability of the two products to accurately identify the presence of snow at specific locations.
.
    \item \textit{Median Difference in Snow Peak per Year (days)}: The median absolute difference, in days, between the predicted and observed timing of the annual snow peak. This metric provides insight into how well the model captures the timing of maximum snow accumulation.

\end{itemize}

\textbf{B. Qualitative Comparison of Spatial Maps}: In addition to the quantitative station-level comparison, we also present a qualitative comparison of the LSTM-derived $'swe'$ maps and ERA5 $'sd'$ maps. This comparison is intended to demonstrate the LSTM's ability to generate spatially consistent and seasonally reasonable snow maps.

\section{Results} \label{sec:results}
\begin{table}[h]
\caption{Comparing the performance of the LSTM classification outputs to the thresholded ERA5 'sd' variable on $100$ batches from the training and testing period.}
\label{tab:res_snowPresence_perf}
\begin{tabular}{|c|ccc|ccc|}
\hline
 & \multicolumn{3}{c|}{\textbf{ERA5}} & \multicolumn{3}{c|}{\textbf{LSTM}} \\ \hline
 & \multicolumn{1}{c|}{\textit{\textbf{Precision}}} & \multicolumn{1}{c|}{\textit{\textbf{Recall}}} & \textit{\textbf{Accuracy}} & \multicolumn{1}{c|}{\textit{\textbf{Precision}}} & \multicolumn{1}{c|}{\textit{\textbf{Recall}}} & \textit{\textbf{Accuracy}} \\ \hline
\begin{tabular}[c]{@{}c@{}}Training Period\\ $01/01/2010 - 12/31/2019$\end{tabular} & \multicolumn{1}{c|}{0.891} & \multicolumn{1}{c|}{0.868} & 0.883 & \multicolumn{1}{c|}{0.948} & \multicolumn{1}{c|}{0.928} & 0.940 \\ \hline
\begin{tabular}[c]{@{}c@{}}Testing Period\\ $01/01/2021-12/31/2021$\end{tabular} & \multicolumn{1}{c|}{0.906} & \multicolumn{1}{c|}{0.876} & 0.894 & \multicolumn{1}{c|}{0.952} & \multicolumn{1}{c|}{0.945} & 0.949 \\ \hline
\end{tabular}
\end{table}

\begin{figure}[h]
    \centering
    \includegraphics[width=\textwidth]{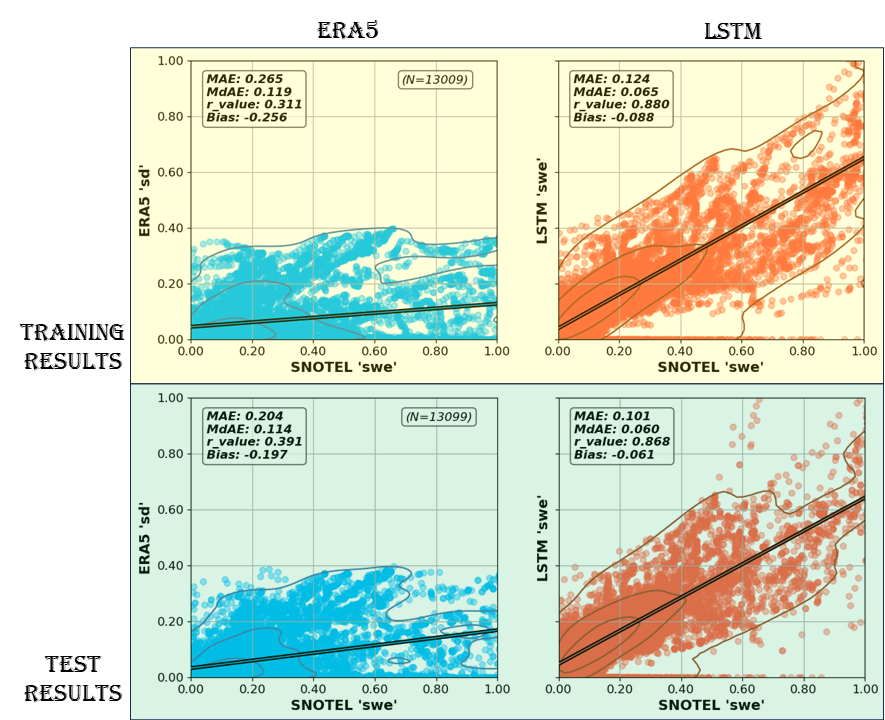} % Change this to your image file name (without extension)
    \caption{Comparison of 'snow-depth' estimates from the ERA5 reanalysis products ($'sd'$) to the LSTM estimated products ($'swe'$).}
    \label{fig:era5_lstm_perfComp}
\end{figure}

\begin{figure}[h]
    \centering
    \includegraphics[width=\textwidth, height=5.2in]{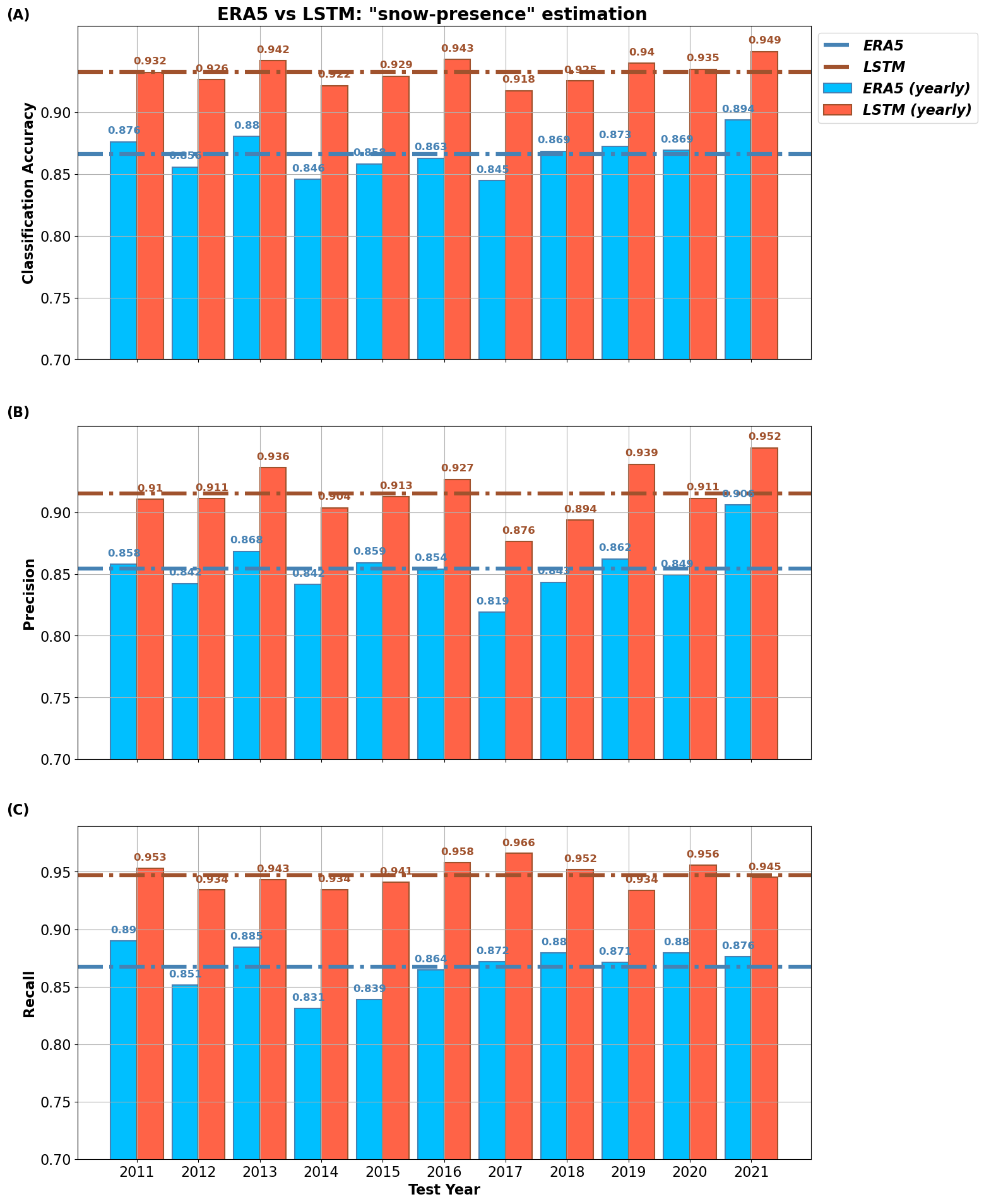} % Change this to your image file name (without extension)
    \caption{Performance of the ERA5 and LSTM products in estimating the 'snow-presence' indicating the performance range of these model when applied to yearly data in terms of (A) Accuracy, (B) Precision, and (c) Recall.}
    \label{fig:era5_lstm_tempCrossVal_class}
\end{figure}

\subsection{LSTM vs ERA5 on a temporal test set} \label{ss_exp1_tempTest_res}
The results of the LSTM and ERA5 models for the temporal test experiment described in Sec. \ref{ss_exp1_tempTest} are presented here, covering both the training and test periods. The ‘snow-presence’ (classification) results are shown in Table \ref{tab:res_snowPresence_perf}, while the ‘snow-depth’ (regression) results are illustrated in Fig. \ref{fig:era5_lstm_perfComp}. In terms of ‘snow-presence’ classification, both the ERA5 and LSTM models perform well, achieving a classification accuracy of $\geq 0.9$. However, the LSTM clearly outperforms the ERA5 estimates by approximately $5\%$. It is worth noting that both methods exhibit slightly higher precision than recall, indicating that there are more False Negatives (FNs)—days incorrectly predicted as having no snow—than False Positives (FPs), where snow is incorrectly predicted. Despite this trend, the LSTM consistently outperforms the ERA5 model in classification accuracy.

In terms of ‘snow-depth’ (regression) (as shown in Fig. \ref{fig:era5_lstm_perfComp}), we present scatter plots that compare the true and predicted snow values for days in the test period where snow was present. The top row shows results for the training period samples, while the bottom row displays results for the test period samples. Similarly, the left column compares the ERA5 $'sd'$ variable to the corresponding SNOTEL $'swe'$ measurements, while the right column compares the LSTM $'swe'$ to the corresponding SNOTEL $'swe'$ measurements. Overall, the LSTM results show much clearer agreement between the true and estimated snow values. In particular, for days with significant snowfall (i.e., in situ measurements of snow $\geq 0.3m$), the ERA5 model tends to significantly underestimate the amount of snow. These results indicate that the LSTM model outperforms ERA5 in both ‘snow-presence’ (classification) and ‘snow-depth’ (regression) tasks. The consistent improvement observed in both Mean Absolute Error (MAE) and Median Absolute Error (MdAE) suggests that the LSTM model enhances point predictions, rather than just improving performance on isolated outliers. Furthermore, the training and testing performance for both the classification and regression models are highly comparable, indicating that the LSTM model is well-trained and generalizes effectively to unseen data.

\begin{figure}[h]
    \centering
    \includegraphics[width=\textwidth, height=4.0in]{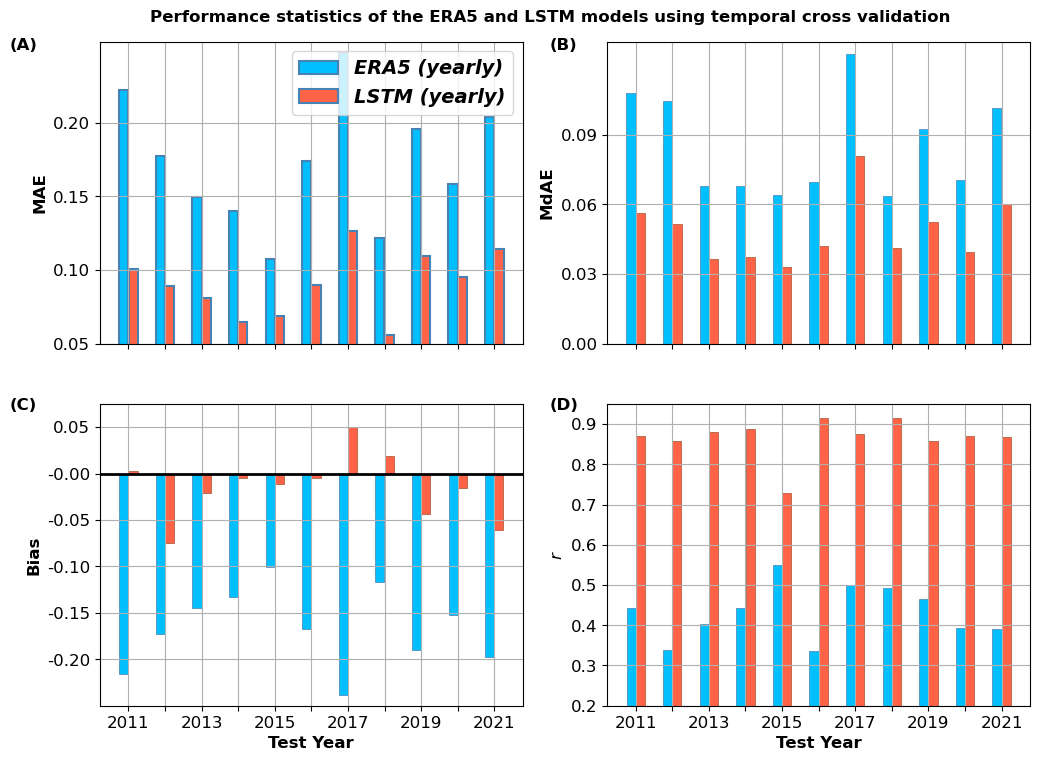} % Change this to your image file name (without extension)
    \caption{Performance of the ERA5 and LSTM products in estimating the 'snow-depth' indicating the performance range of these model when applied to yearly data in terms of (A) MAE, (B) MdAE, (C) Bias, and (D) Pearson-r}
    \label{fig:era5_lstm_tempCrossVal_reg}
\end{figure}

\subsection{Temporal Cross validation results} \label{ss_exp2_tempTest_res}
While the results in the previous section show the ability of the LSTM to perform on previously unseen data from a limited temporal range, it is also essential verify that the improved performance can be repeated across different temporal ranges. The results of temporal cross validation (originally described in Sec. \ref{ss_exp2_tempCV}) for 'snow-presence' (classification) and 'snow-depth' (regression) are shown in Fig. \ref{fig:era5_lstm_tempCrossVal_class} \& \ref{fig:era5_lstm_tempCrossVal_reg} respectively. Additionally, Table \ref{tab:temp_cv_perf_stats} presents the statistics of the various evaluation metrics across the temporal splits.

In the 'snow-presence' (classification) tests, the classification accuracy results (Fig. \ref{fig:era5_lstm_tempCrossVal_class} (A)) indicate that the LSTM consistently outperforms ERA5 across all three metrics. An interesting observation regarding precision and recall is that, as a general trend, recall tends to be higher than precision for most years—except for $2019$ and $2021$ in the case of the LSTM, and $2014$, $2015$, and $2021$ for ERA5. This suggests that both models are generally more prone to False Positives (FPs) than False Negatives (FNs). Similarly, in the 'snow-depth' (regression) analysis, the LSTM consistently outperforms ERA5 across all metrics. A key observation regarding the bias metric (Fig. \ref{fig:era5_lstm_tempCrossVal_reg} (C)) is that, while the LSTM exhibits a lower overall bias, its distribution appears more random compared to ERA5, which consistently shows a negative bias. Regarding outliers, the year 2017 stands out as a clear outlier in terms of MAE, MdAE, and Bias, whereas the only outlier in terms of the Pearson correlation coefficient is $2015$. Additionally, Table \ref{tab:temp_cv_perf_stats} reveals that the yearly statistics for the LSTM exhibit a wider spread, both in terms of the range between the maximum and minimum values and the overall standard deviation.

% \usepackage{multirow}
\begin{table}[]
\caption{The statistics for the various classification and regression metrics when applied to across $11$ yearly cross-validation splits}
\label{tab:temp_cv_perf_stats}
\begin{tabular}{|c|c|ccc|cccc|}
\hline
 &  & \multicolumn{3}{c|}{\textit{\textbf{Classification}}} & \multicolumn{4}{c|}{\textit{\textbf{Regression}}} \\ \hline
 &  & \multicolumn{1}{c|}{\textbf{Accuracy}} & \multicolumn{1}{c|}{\textbf{Precision}} & \textbf{Recall} & \multicolumn{1}{c|}{\textbf{MAE}} & \multicolumn{1}{c|}{\textbf{MdAE}} & \multicolumn{1}{c|}{\textbf{Bias}} & \textbf{$r$} \\ \hline
\multirow{4}{*}{\textit{\textbf{ERA5}}} & \textit{min} & \multicolumn{1}{c|}{0.845} & \multicolumn{1}{c|}{0.819} & 0.831 & \multicolumn{1}{c|}{0.107} & \multicolumn{1}{c|}{0.056} & \multicolumn{1}{c|}{-0.239} & 0.337 \\ \cline{2-9} 
 & \textit{max} & \multicolumn{1}{c|}{0.894} & \multicolumn{1}{c|}{0.906} & 0.890 & \multicolumn{1}{c|}{0.247} & \multicolumn{1}{c|}{0.127} & \multicolumn{1}{c|}{-0.100} & 0.548 \\ \cline{2-9} 
 & \textit{$\mu_{sd}^{ERA5}$} & \multicolumn{1}{c|}{0.866} & \multicolumn{1}{c|}{0.855} & 0.867 & \multicolumn{1}{c|}{0.172} & \multicolumn{1}{c|}{0.090} & \multicolumn{1}{c|}{-0.166} & 0.432 \\ \cline{2-9} 
 & \textit{$\sigma_{sd}^{ERA5}$} & \multicolumn{1}{c|}{0.014} & \multicolumn{1}{c|}{0.021} & 0.018 & \multicolumn{1}{c|}{0.041} & \multicolumn{1}{c|}{0.021} & \multicolumn{1}{c|}{0.040} & 0.064 \\ \hline
\multirow{4}{*}{\textit{\textbf{LSTM}}} & \textit{min} & \multicolumn{1}{c|}{0.918} & \multicolumn{1}{c|}{0.876} & 0.934 & \multicolumn{1}{c|}{0.064} & \multicolumn{1}{c|}{0.033} & \multicolumn{1}{c|}{-0.075} & 0.729 \\ \cline{2-9} 
 & \textit{max} & \multicolumn{1}{c|}{0.949} & \multicolumn{1}{c|}{0.952} & 0.966 & \multicolumn{1}{c|}{0.125} & \multicolumn{1}{c|}{0.081} & \multicolumn{1}{c|}{0.051} & 0.915 \\ \cline{2-9} 
 & \textit{$\mu_{sd}^{LSTM}$} & \multicolumn{1}{c|}{0.933} & \multicolumn{1}{c|}{0.916} & 0.947 & \multicolumn{1}{c|}{0.085} & \multicolumn{1}{c|}{0.048} & \multicolumn{1}{c|}{-0.015} & 0.866 \\ \cline{2-9} 
 & \textit{$\sigma_{sd}^{LSTM}$} & \multicolumn{1}{c|}{0.009} & \multicolumn{1}{c|}{0.021} & 0.010 & \multicolumn{1}{c|}{0.021} & \multicolumn{1}{c|}{0.013} & \multicolumn{1}{c|}{0.034} & 0.047 \\ \hline
\end{tabular}
\end{table}

\begin{figure}[h]
    \centering
    \includegraphics[width=\textwidth, height=5.2in]{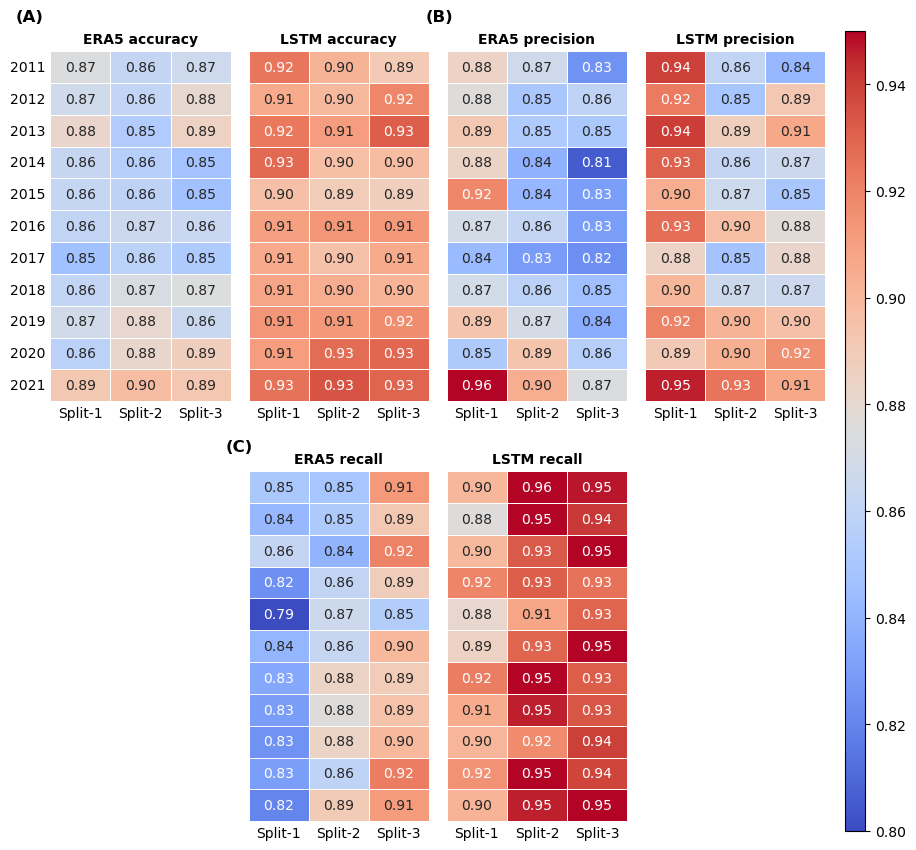} % Change this to your image file name (without extension)
    \caption{Performance of the ERA5 and LSTM products in estimating the 'snow-presence' indicating the performance range of these model when applied to the $33$ spatio-temporal splits in terms of (A) Accuracy, (B) Precision, and (c) Recall.}
    \label{fig:era5_lstm_spatTempCrossVal_class}
\end{figure}

\begin{figure}[h]
    \centering
    \includegraphics[width=\textwidth, height=4.5in]{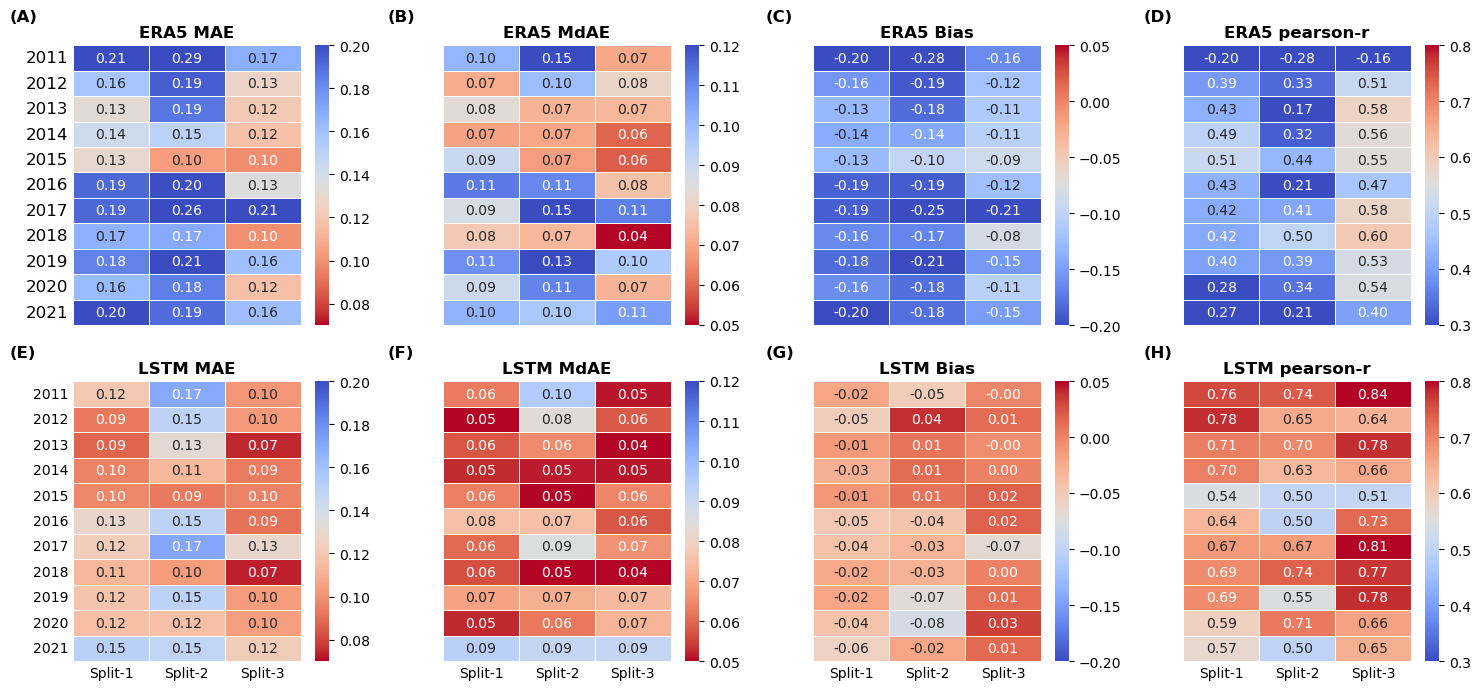} % Change this to your image file name (without extension)
    \caption{Performance of the ERA5 and LSTM products in estimating the 'snow-depth' indicating the performance range of when applied to the $33$ spatio-temporal splits in terms of the different datasets (row) and different metrics (columns). Specifically (A) ERA5 MAE, (B) ERA5 MdAE, (C) ERA5 Bias, (D) ERA5 pearson-r, (E) LSTM MAE, (F) LSTM MdAE, (G) LSTM Bias, and (H) LSTM pearson-r.}
    \label{fig:era5_lstm_spatTempCrossVal_reg}
\end{figure}

% \usepackage{multirow}
\begin{table}[]
\caption{The statistics for the various classification and regression metrics when applied to across $33$ spatiotemporal cross-validation splits}
\label{tab:spatTemp_cv_perf_stats}
\begin{tabular}{|c|c|ccc|cccc|}
\hline
 &  & \multicolumn{3}{c|}{\textit{\textbf{Classification}}} & \multicolumn{4}{c|}{\textit{\textbf{Regression}}} \\ \hline
 &  & \multicolumn{1}{c|}{\textbf{Accuracy}} & \multicolumn{1}{c|}{\textbf{Precision}} & \textbf{Recall} & \multicolumn{1}{c|}{\textbf{MAE}} & \multicolumn{1}{c|}{\textbf{MdAE}} & \multicolumn{1}{c|}{\textbf{Bias}} & \textbf{$r$} \\ \hline
\multirow{4}{*}{\textit{\textbf{ERA5}}} & \textit{min} & \multicolumn{1}{c|}{0.847} & \multicolumn{1}{c|}{0.806} & 0.790 & \multicolumn{1}{c|}{0.099} & \multicolumn{1}{c|}{0.042} & \multicolumn{1}{c|}{-0.280} & -0.280 \\ \cline{2-9} 
 & \textit{max} & \multicolumn{1}{c|}{0.897} & \multicolumn{1}{c|}{0.955} & 0.922 & \multicolumn{1}{c|}{0.285} & \multicolumn{1}{c|}{0.153} & \multicolumn{1}{c|}{-0.084} & 0.603 \\ \cline{2-9} 
 & \textit{$\mu_{sd}^{ERA5}$} & \multicolumn{1}{c|}{0.868} & \multicolumn{1}{c|}{0.862} & 0.265 & \multicolumn{1}{c|}{0.167} & \multicolumn{1}{c|}{0.090} & \multicolumn{1}{c|}{-0.161} & 0.366 \\ \cline{2-9} 
 & \textit{$\sigma_{sd}^{ERA5}$} & \multicolumn{1}{c|}{0.014} & \multicolumn{1}{c|}{0.030} & 0.032 & \multicolumn{1}{c|}{0.043} & \multicolumn{1}{c|}{0.024} & \multicolumn{1}{c|}{0.044} & 0.214 \\ \hline
\multirow{4}{*}{\textit{\textbf{LSTM}}} & \textit{min} & \multicolumn{1}{c|}{0.888} & \multicolumn{1}{c|}{0.842} & 0.876 & \multicolumn{1}{c|}{0.073} & \multicolumn{1}{c|}{0.037} & \multicolumn{1}{c|}{-0.083} & 0.496 \\ \cline{2-9} 
 & \textit{max} & \multicolumn{1}{c|}{0.934} & \multicolumn{1}{c|}{0.946} & 0.956 & \multicolumn{1}{c|}{0.171} & \multicolumn{1}{c|}{0.095} & \multicolumn{1}{c|}{0.039} & 0.842 \\ \cline{2-9} 
 & \textit{$\mu_{sd}^{LSTM}$} & \multicolumn{1}{c|}{0.911} & \multicolumn{1}{c|}{0.893} & 0.927 & \multicolumn{1}{c|}{0.116} & \multicolumn{1}{c|}{0.064} & \multicolumn{1}{c|}{-0.018} & 0.669 \\ \cline{2-9} 
 & \textit{$\sigma_{sd}^{LSTM}$} & \multicolumn{1}{c|}{0.013} & \multicolumn{1}{c|}{0.029} & 0.025 & \multicolumn{1}{c|}{0.021} & \multicolumn{1}{c|}{0.015} & \multicolumn{1}{c|}{0.031} & 0.094 \\ \hline
\end{tabular}
\end{table}

\subsection{Spatio-Temporal Cross validation results} \label{ss_exp2_spatTempTest_res}
While the previous experiment clearly demonstrates the LSTM's ability to generalize across a wide range of temporal test sets, a previously unexplored aspect is its ability to generalize to previously unseen spatial regions. To address this, this section presents the results of the LSTM when trained and evaluated on spatio-temporally discrete test sets (the full experiment settings are described in Sec. \ref{ss_exp3_spatTempCV}). The performance results for 'snow-presence' (classification) and 'snow-depth' (regression) are shown in Fig. \ref{fig:era5_lstm_spatTempCrossVal_class} and Fig. \ref{fig:era5_lstm_spatTempCrossVal_reg}, respectively. Additionally, Table \ref{tab:spatTemp_cv_perf_stats} presents the statistics for the various evaluation metrics across the temporal splits.

In the 'snow-presence' (classification) tests, the classification accuracy results (Fig. \ref{fig:era5_lstm_spatTempCrossVal_class} (A)) indicate that the LSTM consistently outperforms ERA5 across all spatiotemporal splits. However, it is also notable that performance on the spatio-temporal splits is slightly lower than on the yearly splits presented in the previous section. An noteworthy point regarding precision and recall is that, for the LSTM, spatial Split-1 exhibits higher precision and lower recall compared to spatial Split-3. While the overall trend of higher recall than precision for the LSTM continues, spatial Split-1 stands out as an outlier, consistently showing higher precision than recall (interestingly, ERA5 results exhibit the same trend). In the 'snow-depth' (regression) results (Fig. \ref{fig:era5_lstm_spatTempCrossVal_reg}), it is important to note that MAE and MdAE are designed such that lower values are better. Unlike the other metrics, we have inverted the color scale (with low values appearing red and high values appearing blue) for better interpretability. Once again, the LSTM consistently outperforms ERA5 across all temporal splits. However, similar to the classification results, there is a slight drop in performance on the spatio-temporal splits compared to the purely temporal splits. Another notable observation is that the standard deviation among the metrics increases for the spatial cross-validation tests relative to the temporal cross-validation tests—except for regression bias, which remains approximately the same. 

\subsection{ERA5 vs LSTM comparision} \label{ss_exp4_snotelStatPerf_res}
\subsubsection*{A. Station-level comparison}
\begin{figure}[h]
    \centering
    \includegraphics[width=0.7\textwidth, height=4.5in]{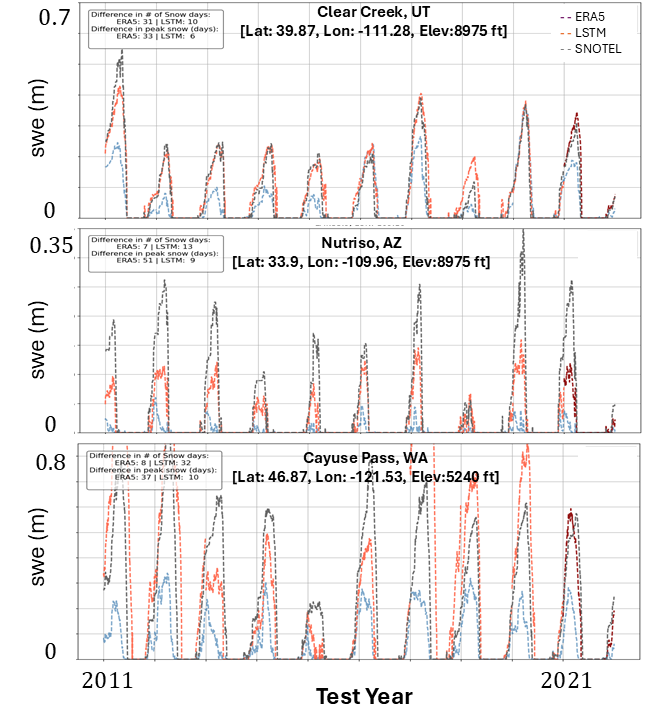} % Change this to your image file name (without extension)
    \caption{Comparison of LSTM and ERA5 estimated snowpack variables to the physical SNOTEL measurements at selected SNOTEL towers}
    \label{fig:snotelStatPerf_res}
\end{figure}

\begin{table}[]
\centering
\caption{Comparison of ERA5 vs LSTM estimates in terms of 'snow dynamics'}
\label{tab:snotelStat_perfStats}
\begin{tabular}{|c|l|l|}
\hline
\multicolumn{1}{|l|}{\textbf{}}                                                                             & \textbf{ERA5} & \textbf{LSTM} \\ \hline
\textit{\textbf{\begin{tabular}[c]{@{}c@{}}Median Difference in Snow \\ Days per year (days)\end{tabular}}} & 44.23         & 31.43         \\ \hline
\textit{\textbf{\begin{tabular}[c]{@{}c@{}}Median Difference in Snow peak\\ each year (days)\end{tabular}}} & 23.94         & 19.25         \\ \hline
\end{tabular}
\end{table}

While the different experiments in Secs.~\ref{ss_exp1_tempTest_res}--\ref{ss_exp2_spatTempTest_res} demonstrate the overall performance of the models—both in terms of estimation accuracy and generalization to previously unseen data—they do not provide direct evidence of the LSTM's ability to capture the temporal dynamics of snowpack evolution. This experiment aims to address that gap by evaluating the LSTM’s ability to replicate changes in the snowpack at individual SNOTEL stations.

Figure~\ref{fig:snotelStatPerf_res} illustrates the variation between the predicted and observed snowpack dynamics at selected SNOTEL stations. Notably, across most sites, the LSTM-derived snow estimates more closely track the actual snowpack changes compared to ERA5.

Table~\ref{tab:snotelStat_perfStats} summarizes the performance of both ERA5 and LSTM across all SNOTEL stations\footnote{This analysis only includes stations with complete data coverage for the entire period.}.

\subsubsection*{B. Qualitative Comparison of Spatial Maps}
\begin{figure}[h]
    \centering
    \includegraphics[width=0.7\textwidth, height=4.5in]{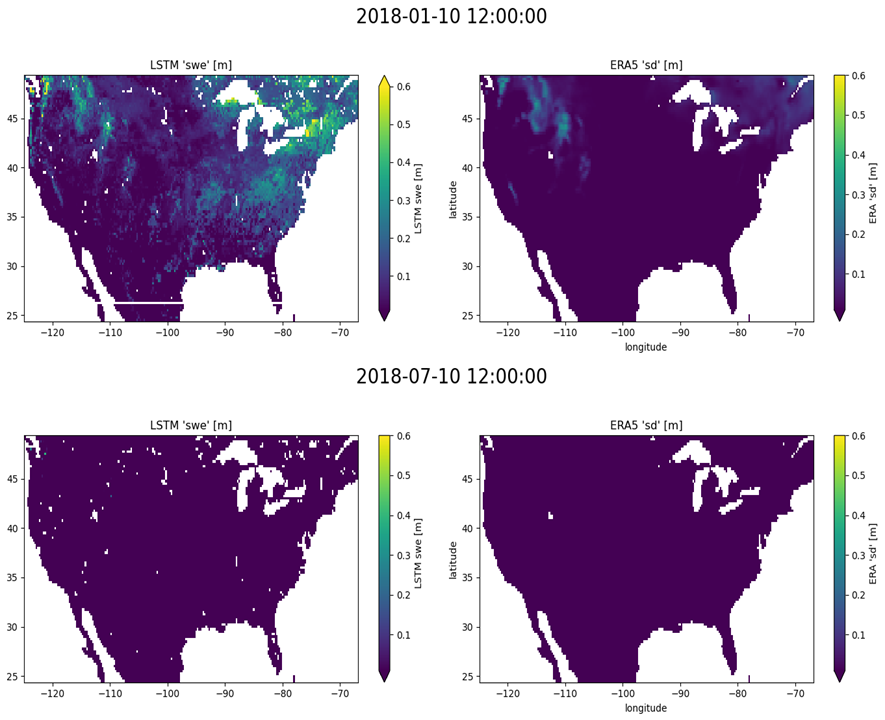} % Change this to your image file name (without extension)
    \caption{Comparison of LSTM and ERA5 estimated snowmaps for CONUS in the winter (top row) and summer (bottom row) of $2018$}
    \label{fig:conus_snowmap_comp}
\end{figure}

Fig. \ref{fig:conus_snowmap_comp} shows the comparison of the LSTM and ERA5 snow maps for CONUS in the winter and summer of $2018$. Both products are able to meet basic "common-sense" requirements of no snow in the summer and increased snow activity in the winter. Additionally, it is known that the East coast of the United States was impacted by a historic "bomb cylone" between January $3-5,~ 2018$ \footnote{\url{https://en.wikipedia.org/wiki/January_2018_North_American_blizzard}},with snow reported as far south as Georgia and Florida. The LSTM snow map from January $10,~ 2018$  accurately captures this event and records heightend snow in the East coast of the US. On the other hand the ERA5 model only represent snow in the North East of the US which is consistent with expected trends of snow presence rather than the effects of some rare events.

\section{Discussion} \label{sec:disc}
The main contribution of this paper is the development of a meteorologically, physically, and morphologically driven LSTM model to predict various snow-pack variables. The results in Sec. \ref{ss_exp1_tempTest_res} clearly demonstrate the improved ability of the LSTM model to predict both components of the snow-pack: 'snow-presence' (classification) and 'snow-depth' (regression). In terms of 'snow-presence', it should be noted that both models achieve high accuracy, with classification accuracies of approximately $90\%$. The ERA5 model achieves an accuracy of $89.10\%$, with a precision of $90.11\%$ and a recall of $87.58\%$. In comparison, the LSTM model achieves a significantly higher accuracy of $94.49\%$, with a precision of $94.99\%$ and a recall of $93.98\%$. A notable observation is that both models exhibit higher precision than recall for the year $2021$, indicating that both the ERA5 reanalysis products and the SNOTEL-trained LSTM model are more likely to miss snow days (false negatives) rather than incorrectly classify a snow-free day as a snow day (false positives). Additionally, the LSTM 'snow-presence' models perform consistently well on both the training and test sets—in fact, they show slightly better performance on the test set. This suggests that the models do not suffer significantly from overfitting (see Table \ref{tab:res_snowPresence_perf} for a detailed performance breakdown). For 'snow-depth' (regression), the LSTM model demonstrates a significant improvement over the ERA5 reanalysis products across multiple performance metrics. On both the training and test sets, the LSTM model outperform ERA5 $'sd'$ estimates by approximately $100\%$. Specifically, for the test set, the LSTM model's per-sample error (MAE: $0.101$, MdAE: $0.06$) is significantly lower than the corresponding ERA5 $'sd'$ estimates (MAE: $0.204$,  MdAE: $0.114$). Additionally, it should also be noted that the LSTM model's predictions show much stronger correlation with true values, as illustrated in the scatter plots in Fig. \ref{fig:era5_lstm_perfComp}. This improvement is also reflected in the Pearson correlation coefficients, where the LSTM model (Pearson-r: $0.868$) significantly outperforms the ERA5 estimates (Pearson-r: $0.391$). Another key observation is that both the ERA5 and LSTM models exhibit a negative bias for the year $2021$, meaning they tend to underestimate 'snow-presence'. Further, the LSTM model’s bias ($-0.061$) is significantly smaller compared to the ERA5 product ($-0.197$). These findings highlight the LSTM model’s superior performance in both classification and regression tasks, making it a more reliable tool for snow pack prediction compared to ERA5 reanalysis products.

In addition to the temporal test sets, we also conducted temporal cross-validation, as reported in Sec. \ref{ss_exp2_tempTest_res}. These results demonstrate that the LSTM model’s performance remains consistent across the entire temporal range of the available dataset. The main takeaway from both the 'snow-presence' (classification) and 'snow-depth' (regression) experiments is that the LSTM models generalize effectively across a wide temporal test range. An interesting observation is that over longer temporal test ranges, the recall is consistently (and on average) higher than precision. This suggests that, except for a few outliers (such as $2021$), both the LSTM and ERA5 models are more prone to generating false positives (FPs) than false negatives (FNs). However, this does not follow a strict trend; rather, there appears to be a clear temporal dependence on this behavior. In terms of the 'snow-depth', an interesting observation is that the the LSTM consistently improves all the metrics. The main outlier in terms of the performance is the year $2017$, as the LSTM models show the poorest performance on the per sample metrics - namely MAE, MdAE, and Bias. Despite this, the Pearson correlation coefficient (Pearson-r) remains high for 2017. This suggests that while the LSTM model produces higher errors for some samples (outliers), the overall trend in predictions remains consistent with observed data. Conversely, the year with the lowest Pearson-r is $2015$, indicating that while the model generally performs well, it struggles with certain high-snow samples. Another key takeaway from Table \ref{tab:temp_cv_perf_stats} is that LSTM predictions exhibit a much smaller spread across all metrics (for both classification and regression tasks) compared to ERA5. Additionally, the LSTM model has a lower standard deviation in performance, highlighting its greater consistency across the entire temporal test range. Furthermore, the significant reduction in bias in the LSTM predictions compared to the ERA5 estimates is particularly noteworthy, reinforcing the LSTM model’s superiority in both accuracy and consistency.

The final experiment tested the LSTM model in a spatio-temporal cross-validation scenario, where the test set had no spatial or temporal overlap with the training set. The results, reported in Sec. \ref{ss_exp2_spatTempTest_res}, confirm the superiority of the LSTM method over the ERA5 reanalysis products and highlight its ability to generalize with high accuracy to previously unseen data across both sub-tasks of snow-pack estimation. However, the results indicate that performance on a fully spatiotemporal test set is, on average, slightly lower than on the temporal test sets (Sec. \ref{ss_exp2_tempTest_res}) for both classification and regression tasks. This decline suggests that variations in spatial properties—such as morphology, land use, or vegetation—may significantly impact model performance. Additionally, while the spatio-temporal cross-validation tests provide valuable insights into how the model generalizes across different climatic, ecological, and geological conditions within the training dataset, they are not explicitly designed to stress-test performance in entirely novel bio-geographical regions not represented in the training data. Understanding these limitations is crucial for identifying potential improvements in LSTM models or other data-driven inference techniques. Nevertheless, the experiments presented here clearly demonstrate that ML-based LSTM models that leverage time-series weather data are highly effective estimators of snow-pack variables, generally surpassing (or at least matching) large-scale weather reanalysis products in terms of both 'snow-presence' (classification) and 'snow-depth' (regression) performance.

\section{Conclusions} \label{sec:conc}

In this manuscript, we have presented an LSTM-based machine learning model with a long lookback period that provides high-quality estimates of snow-pack, specifically Snow Water Equivalent (SWE). The model generates these SWE estimates using a combination of weather and climatological time-series data (primarily sourced from the ERA5 database) and static bio-geographic variables that capture the spatio-morphological characteristics of each location. Our experiments demonstrate that LSTM-based models consistently outperform standard ERA5 reanalysis products in estimating snow-pack variables, particularly SWE. This improved performance holds even when tested on datasets with no temporal (Sec. \ref{ss_exp2_tempTest_res}) or spatio-temporal (Sec. \ref{ss_exp2_spatTempTest_res}) overlap with the training data. These results indicate that the LSTM model is capable of generalizing effectively to previously unseen data, making it a robust tool for snow-pack estimation.

While the experiments presented in this paper clearly demonstrate the ability of the LSTM-based model to accurately capture snow-pack behavior and generalize to previously unseen data, multiple avenues exist for improving model performance and enhancing our understanding of these models. The first avenue we are exploring is testing the ability of LSTM models to generalize more effectively when applied to previously unseen bio-geographic regimes. Another area of improvement under investigation is hyperparameter tuning, aimed at identifying the optimal architecture and parameter combinations for large-scale applications. Additionally, we are examining the potential of incorporating snowfall and snow depth data from alternative sources (such as \url{https://www.ncei.noaa.gov/access/monitoring/daily-snow/}) or reprocessed SWE datasets (e.g., those presented in \cite{fontrodona2023nh}) to further enhance model performance.

\appendix
\renewcommand{\thesection}{Appendix-\Alph{section}:}

\section{LSTM-based fully independent snow-pack estimation} \label{app:lstm_independance}
\begin{table}[h]
\caption{Effect of ERA5 snow variables on the performance of the LSTM on estimating the 'snow-presence'.}
\label{tab:res_snowPresence_perf_noERAsnow}
\begin{tabular}{|c|ccc|ccc|}
\hline
 & \multicolumn{3}{c|}{\textbf{LSTM-est}} & \multicolumn{3}{c|}{\textbf{LSTM}} \\ \hline
 & \multicolumn{1}{c|}{\textit{\textbf{Precision}}} & \multicolumn{1}{c|}{\textit{\textbf{Recall}}} & \textit{\textbf{Accuracy}} & \multicolumn{1}{c|}{\textit{\textbf{Precision}}} & \multicolumn{1}{c|}{\textit{\textbf{Recall}}} & \textit{\textbf{Accuracy}} \\ \hline
\begin{tabular}[c]{@{}c@{}}Training Period\\ $01/01/2010 - 12/31/2019$\end{tabular} & \multicolumn{1}{c|}{0.931} & \multicolumn{1}{c|}{0.962} & {0.946} & \multicolumn{1}{c|}{0.948} & \multicolumn{1}{c|}{0.928} & 0.940 \\ \hline
\begin{tabular}[c]{@{}c@{}}Testing Period\\ $01/01/2021-12/31/2021$\end{tabular} & \multicolumn{1}{c|}{0.938} & \multicolumn{1}{c|}{0.969} & {0.953} & \multicolumn{1}{c|}{0.952} & \multicolumn{1}{c|}{0.945} & 0.949 \\ \hline
\end{tabular}
\end{table}

\begin{figure}[h]
    \centering
    \includegraphics[width=\textwidth]{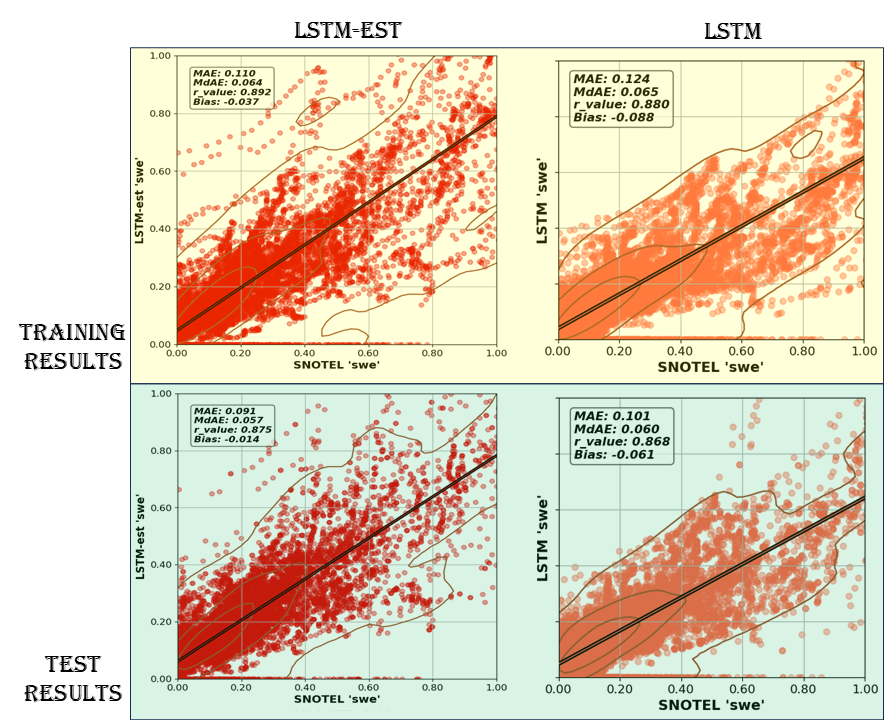} % Change this to your image file name (without extension)
    \caption{Comparison of LSTM estimated products ($'swe'$) products when (a) no ERA5 snow related variables are used ('LSTM-est') as input, versus (b) ERA5 snow-related prodcuts are used as input.}
    \label{fig:lstm_perfComp_noERAsnow}
\end{figure}

While the experiments (Sec.~\ref{subsec:modelTesting}) and results (Sec.~\ref{sec:results}) presented in this manuscript clearly highlight the ability of the LSTM models to effectively capture snowpack dynamics across a broad range of physical and geographical conditions, it is important to note that the models considered in these experiments include several snow-specific input variables that support estimation performance. These snow-specific variables include the ERA5 Snow Depth ($'sd'$), Convective Snowfall Rate water equivalent, ($'csfr'$), Snow Evaporation ($'es'$), and Snow Melt ($'smlt'$). These variables are highlighted in green in Table~\ref{tab:era5_var}, with $'sd'$ additionally marked in red, as it is the primary ERA5 variable used in comparison against the LSTM-estimated $swe$.

Because the LSTM models leverage ERA5-derived snow variables as inputs, their performance can, to some extent, be interpreted as bias correction—where the model adjusts or fine-tunes ERA5-derived snow estimates to better match the \textit{in situ} SNOTEL measurements. While such behavior is useful for generating improved snow products, it does not demonstrate whether the physical and meteorological variables, excluding direct snow proxies, are sufficient to independently estimate the properties of the snowpack. The experiment in this section aims to address that question.

To test whether the model can infer snowpack dynamics solely from a feature set that excludes such proxy variables, we replicate the experiment described in Sec.~\ref{ss_exp1_tempTest}, this time omitting all snow-related ERA5 variables (highlighted in color in Table~\ref{tab:era5_var}). By comparing the performance of this snow-excluded model (referred to here as ‘LSTM-est’) against the original model (results presented in Sec.~\ref{ss_exp1_tempTest_res}), we can evaluate the contribution of snow-specific variables to the model’s predictive ability. This comparison helps assess whether the model is learning meaningful associations between snowpack and the remaining physical or morphological predictors, or simply correcting biases from ERA5 snow inputs.

The performance of the LSTM-est model, relative to the original LSTM, is summarized in Table~\ref{tab:res_snowPresence_perf_noERAsnow} (for snow presence classification) and Fig.~\ref{fig:lstm_perfComp_noERAsnow} (for snow depth regression). Based on these results, there appears to be no significant performance degradation when snow-specific inputs are excluded. Interestingly, the precision and recall metrics for the LSTM-est model show an inverse trend compared to the original LSTM. In terms of regression, the two models show nearly identical performance, though the snow-excluded version yields slightly reduced bias. These results suggest that the remaining input variables carry sufficient signal for the LSTM to effectively learn and estimate snowpack dynamics, even in the absence of direct snow predictors.

\printbibliography
\end{document}